\documentclass{amsart}
\usepackage{graphicx,tikz,bm}
\usepackage{amsmath,amsfonts,amssymb}
\usepackage{color}
\usetikzlibrary{decorations.markings}
\tikzstyle{vertex}=[circle, draw, inner sep=0pt, minimum size=4pt]
\tikzstyle{bvertex}=[circle, draw, fill=black, inner sep=0pt, minimum size=6pt]
\tikzstyle{emptynode}=[circle]

\renewcommand{\ker}{{\mathbf{ker}\,}}
\newcommand{\im}{{\mathbf{im}\,}}
\newcommand{\spn}{{\mathbf{span}\,}}
\newcommand{\dm}{{\mathbf{dim}\,}}

\newtheorem{theorem}{Theorem}[section]
\newtheorem{lemma}[theorem]{Lemma}
\theoremstyle{definition}
\newtheorem{definition}[theorem]{Definition}
\newtheorem{problem}[theorem]{Problem}
\newtheorem{example}[theorem]{Example}
\newtheorem{corollary}[theorem]{Corollary}

\theoremstyle{remark}

\numberwithin{equation}{section}

\begin{document}
\title{Kron Reduction of \\Generalized Electrical Networks} 

\author{Sina Y. Caliskan}
\address{ Department of Electrical Engineering, University of California at Los Angeles, CA 90095-1594, United States}
\email{caliskan@ee.ucla.edu}

\author{Paulo Tabuada}
\address{ Department of Electrical Engineering, University of California at Los Angeles, CA 90095-1594, United States}
\email{tabuada@ee.ucla.edu}


\date{}

\begin{abstract}                        
Kron reduction is used to simplify the analysis of multi-machine power systems under certain steady state assumptions that underly the usage of phasors. Using ideas from behavioral system theory, we show how to perform Kron reduction for a class of electrical networks without steady state assumptions. The reduced models can thus be used to analyze the transient as well as the steady state behavior of these electrical networks.
\end{abstract}

\maketitle

\section{Introduction}

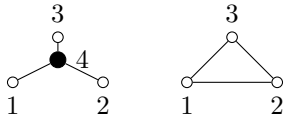
\begin{figure}[t]
\begin{center}
\begin{tabular}{ccc}
\begin{tikzpicture}[scale=0.6]
    \node[vertex] (1) at (1,0) [label=below:$1$]{};
    \node[bvertex] (4) at (2,0.5) [label=right:$4$]{};
    \node[vertex] (3) at (2,1) [label=above:$3$]{};
    \node[vertex] (2) at (3,0) [label=below:$2$]{};
    \path
    (1) edge (4)
    (2) edge (4)
    (4) edge (3);
\end{tikzpicture}
&  &
\begin{tikzpicture}[scale=0.6]
    \node[vertex] (1) at (1,0) [label=below:$1$]{};
    \node[vertex] (3) at (2,1) [label=above:$3$]{};
    \node[vertex] (2) at (3,0) [label=below:$2$]{};
    \path
    (1) edge (2)
    (2) edge (3)
    (1) edge (3);
\end{tikzpicture}\\
\end{tabular}
\end{center}
\label{Y-D}
\caption{Equivalent $Y$ (left) and $\Delta$ (right) circuits. The internal vertex depicted in black is eliminated in the conversion from the $Y$ to the $\Delta$ circuit.}
\end{figure}

Multi-machine power networks are the interconnection of power
generators and substations via three-phase transmission lines. This structure can be abstracted as a graph in which edges represent transmission lines and vertices represent buses that could be connected to generators and/or to substations abstracted as loads. Depending on the models used to describe the generators and the loads, we have a set of algebraic, differential, or
algebro-differential equations per vertex. When the number of vertices
increases this set of equations quickly becomes intractable.
Common practice in the power systems literature is to reduce this
set of equations, through a process called Kron reduction
\cite[Sec. 9.3]{BergenBook},
that results in a simpler set of equations providing the same
relationships between voltage and current at the generators'
terminals. One example of Kron reduction is the classical $Y$ to $\Delta$ conversion depicted in Figure~\ref{Y-D}. 
We regard the white vertices as boundary vertices to which generators are connected. The black vertex is an internal vertex that is connected to a load. The $Y$ to $\Delta$ conversion provides an equivalent circuit solely consisting of boundary vertices. 

Despite its widespread use, Kron reduction is based on the use of
\emph{phasors} and it requires the current and voltage waveforms
in each phase to be sinusoidal and with the same frequency. This
assumption seems contradictory if we want to study the transient
behavior of a power system during  which the waveforms are
\textit{not} sinusoidal.

The contribution of this paper is to identify a class of electric
networks for which Kron reduction can be performed in the time
domain, \textit{i.e.}, without resorting to phasors. The reduced
models can then be used to study the transient as well as the
steady state behavior of these electrical networks which can be
used to describe, \textit{e.g.}, short transmission lines.

A graph-theoretic discussion of Kron reduction can be found
in~\cite{Dorfler2012}. Our results are based on the very same
graph-theoretic constructions. The problem of Kron reduction can
be understood as a system equivalence problem: when do two models
- the original and the Kron reduced - describe the same terminal
behavior? Here we use the term behavior in the sense of behavioral systems theory~\cite{WillemsBook}. This problem was first solved for purely resistive
circuits in~ \cite{Willems2009} and then for RLC circuits
in~\cite{Verriest2010} by describing not only when two models are
equivalent but also by characterizing the set of all such equivalent models. In the
context of Kron reduction we are interested in constructing the
smallest model that describes the relationships between voltage
and current at the generators' terminals. This problem was solved
for purely resistive, purely inductive and purely capacitive
circuits in~\cite{Arjan2009,Arjan2010} and for a class of RL
circuits called \textit{homogeneous} RL
networks in~\cite{Caliskan2012}. This paper generalizes the results
in~\cite{Caliskan2012} by describing a larger class of electrical
networks for which Kron can be performed in the time domain. 

\section{Notation and Definitions}

Let $\mathbb{N}$ be the set of natural numbers, $\mathbb{R}$ be
the set of real numbers, $\mathbb{R}_{+}$ be the set of all
strictly positive real numbers and $\mathbb{R}_{+0} = \mathbb{R}_+
\cup \{0\}$. For any $n \in \mathbb{N}$, the set $\bar{n}$ is
defined as $\bar{n} = \{1,\ldots,n\}$. The cardinality of the set
$S$ is denoted by $|S|$. We also use $|x|$ to denote the absolute
value of $x \in \mathbb{R}$. The set of all real $n\times m$ matrices
 is denoted by $\mathbb{R}^{n\times m}$. The
element of $\mathbb{R}^n$ with all entries equal to 0 is
denoted by $\bm{0}_n$. We also use
$\bm{0}_n$ to denote the zero map \mbox{$\bm{0}_n:\mathbb{R}
\rightarrow \mathbb{R}^n$}, which maps every real number to
$\bm{0}_n \in \mathbb{R}^n$. For any matrix $M \in
\mathbb{R}^{n\times m}$, we denote the element located in the
$i^{\text{th}}$ row and the $j^{\text{th}}$ column by $M_{ij}$.
For any set of vectors $S =
\{v_1,\ldots,v_n\}$, $\spn S$ is the vector space spanned by the
elements of $S$. For any vector space $V$, $\dm V$ denotes the
dimension of $V$. The kernel and the image of a linear map $f$ are
denoted by $\ker f$ and $\im f$. The set of all smooth functions
with domain $A$ and codomain $B$ is denoted by
$\mathcal{C}^\infty(A,B)$.

We follow \cite{GodsilBook} for the definitions related with
graphs. A \textit{graph} $\mathcal{G}=(\mathcal{V},\mathcal{E})$
is a two-tuple where $\mathcal{V}$ is a finite set of vertices and $\mathcal{E} \subseteq \mathcal{V} \times
\mathcal{V}$ is a finite set of edges. If the edges have no sense of direction, the graph
is called an \textit{undirected graph}. In directed graphs, $(x,y)
\in \mathcal{E}$ implies that vertices $x$ and $y$ are connected via
an edge with tail vertex $x$ and head vertex $y$. Let $v =
|\mathcal{V}|$ and $e = |\mathcal{E}|$. Without loss of
generality, we will assume $\mathcal{V} = \bar{v}$ for the rest of
the paper. Any graph $\mathcal{G}$ is completely represented by a
$v \times e$ matrix called \textit{incidence matrix} $B$. The rows
of the incidence matrix represent the vertices and the columns
represent the edges. In directed graphs, every edge $e_k = (i,j)$,
$k \in \bar{e}$ is encoded in the incidence matrix by setting $B_{ik} = -1$,
$B_{jk} = 1$ and $B_{xk} = 0$ for all $x \in
\bar{v} \backslash \{i,j\}$. The set of adjacent vertices
$\mathcal{A}$ is defined as \mbox{$\mathcal{A} = \{(x,y) \in
\mathcal{V} \times \mathcal{V}~:~(x,y)\in
\mathcal{E}~\text{or}~(y,x) \in \mathcal{E}\}$}. A \textit{path}
of length $\ell$ from vertex $i$ to vertex $j$ is the subset of
vertices $\{r_0,\ldots,r_\ell\}$ such that $r_0 = i$, $r_\ell =
j$, and $(r_{k-1},r_{k})\in \mathcal{A}$ for all $k \in
\bar{\ell}$. The careful reader may notice that we allow
traversing directed edges in both directions. A graph is
called a \textit{connected graph} if for every $i,j \in
\mathcal{V}$, $i \ne j$, there exists a path from $i$ to $j$. A
\textit{cycle} of a graph is a connected subgraph in which every
vertex has exactly two neighbors. For the rest of the paper, we
will assume that every graph is connected.

We will need the following lemma, which is a restatement of part of Theorem 3.1 in \cite{Arjan2010}, to prove our main result. A similar result can also be found in Lemma 2.1 in \cite{Dorfler2012}.

\begin{lemma}[Theorem 3.1 in \cite{Arjan2010}]
\label{lemmakron} Let $\mathcal{G} = (\mathcal{V},\mathcal{E})$ be
a directed and connected graph with nontrivial subsets
$\mathcal{V}_i \subset \mathcal{V}$ and $\mathcal{V}_b = \mathcal{V} \backslash \mathcal{V}_i \subset \mathcal{V}$. Let $B_b$ and $B_i$ be the matrices that are obtained by collecting the rows of $B$ that correspond to vertices in the sets $\mathcal{V}_b$ and $\mathcal{V}_i$, respectively. Then there exist a unique directed graph
$\hat{\mathcal{G}} = (\mathcal{V}_b,\hat{\mathcal{E}})$ with
incidence matrix $\hat{B}_b \in \{-1,0,1\}^{|\mathcal{V}_b| \times
|\hat{\mathcal{E}}|}$ and a diagonal matrix $W_r \in
\mathbb{R}_+^{|\hat{\mathcal{E}}| \times |\hat{\mathcal{E}}|}$
such that $$\hat{B}_bW_r\hat{B}^T_b = B_bWB^T_b
- B_bWB^T_i(B_iWB^T_i)^{-1}B_iWB^T_b.$$
\end{lemma}

Note that Lemma \ref{lemmakron} implies that  matrix $B_iWB^T_i$ is invertible for any diagonal matrix $W \in \mathbb{R}^{e \times e}$ with strictly positive diagonal elements, where $B_i$ is defined as in Lemma \ref{lemmakron}.

\section{Kirchoff's Laws on Graphs}

We review Kirchoff's Current Law (KCL) and Kirchoff's Voltage Law
(KVL) on abstract graphs, which were introduced in
\cite{Arjan2009,Arjan2010}. Let $\mathbb{R}^S$ be the vector space
of all functions from $S$ to the field of real numbers
$\mathbb{R}$. The space of trajectories of currents entering into
vertices is denoted by $\Lambda_0$, and it is defined as $\Lambda_0
=\mathcal{C}^\infty (\mathbb{R},\mathbb{R}^\mathcal{V})$. We can think of vertices as zero-dimensional objects and for this reason we use the subscript $0$. We identify $\mathbb{R}^\mathcal{V}$ with the vector space
$\mathbb{R}^v$ and let $(\mathbb{R}^{v})^*$ be the dual space of
$\mathbb{R}^v$. The dual space of $\Lambda_0$ is denoted by
$\Lambda^0$. We can interpret $\Lambda^0 =\mathcal{C}^\infty
(\mathbb{R},(\mathbb{R}^v)^*)$ as the space of trajectories of
voltages on the vertices. Similarly, the space of trajectories of
currents flowing through the edges is denoted by $\Lambda_1$, and
it is defined as $\Lambda_1 = \mathcal{C}^\infty
(\mathbb{R},\mathbb{R}^\mathcal{E})$. Edges are regarded as one dimensional objects, hence the subscript $1$. We identify $\mathbb{R}^\mathcal{E}$ with the vector space $\mathbb{R}^e$ and interpret the dual space of $\Lambda_1$, denoted by $\Lambda^1 =\mathcal{C}^\infty
(\mathbb{R},(\mathbb{R}^e)^*)$, as the
space of trajectories of voltages across the edges. We can think
of the incidence matrix $B$ of the directed graph $\mathcal{G}$ as the matrix
representation of the linear map
\begin{equation}
\label{operB} B~:~\mathbb{R}^e \rightarrow \mathbb{R}^v.
\end{equation}
Using the standard inner product in Euclidean spaces as the
duality product, the dual map of~(\ref{operB}) can be represented by the matrix $B^T$. The linear
map~(\ref{operB}) can be extended to the map
\mbox{$\bar{B}~:~\Lambda_1 \rightarrow \Lambda_0$} defined by
\mbox{$(\bar{B} \circ f_1)(t) = B \circ f_1(t)$} for every $f_1
\in \Lambda_1$ and every $t \in \mathbb{R}$. Similarly, we can
define the dual map \mbox{$\bar{B^T}~:~\Lambda^0 \rightarrow
\Lambda^1$} as \mbox{$(\bar{B^T} \circ g^0)(t) = B^T \circ
g^0(t)$} for every $g^0 \in \Lambda^0$ and every $t \in
\mathbb{R}$. For the graph \mbox{$\mathcal{G} =
(\mathcal{V},\mathcal{E})$}, let $\{\mathcal{V}_b,\mathcal{V}_i\}$
be a partition of its set of vertices $\mathcal{V}$. In other
words, $\mathcal{V} = \mathcal{V}_b \cup \mathcal{V}_i$ and
$\mathcal{V}_b \cap \mathcal{V}_i = \emptyset$. We call the
elements of $\mathcal{V}_b$ \textit{boundary vertices} and the
elements of $\mathcal{V}_i$ \textit{internal vertices}. A graph
$\mathcal{G}$ with at least one boundary vertex is called an
\textit{open graph} \cite{Arjan2009}. We will use open graphs to model power networks connected to generators through its boundary vertices. The partition
$\{\mathcal{V}_b,\mathcal{V}_i\}$ of $\mathcal{V}$ leads to a
decomposition of the incidence matrix
$B$. Explicitly, the incidence matrix can be decomposed as $B =[B_i^T B_b^T]^T.$ In this decomposition, $B_b$ contains the rows
of the incidence matrix that correspond to boundary vertices, and
$B_i$ contains the rows of the incidence matrix that correspond to
internal vertices. The decomposition of $B$ into $B_b$ and $B_i$
induces the decomposition of the vector space $\Lambda_0$ into
$\Lambda_{0b}$ and $\Lambda_{0i}$. Similarly, the dual space
$\Lambda^0$ can be decomposed into $\Lambda^{0b}$ and
$\Lambda^{0i}$. We can define the linear maps
$\bar{B}_b~:~\Lambda_1 \rightarrow \Lambda_{0b}$,
$\bar{B}_i~:~\Lambda_1 \rightarrow \Lambda_{0i}$ and their dual
maps as before.

Kirchoff's Current Law states that the sum of currents entering
into a vertex is zero. We can conveniently describe the sum of the currents entering each vertex in a directed graph by $\bar{B}I_1$, where $I_1 \in \Lambda_1$ is the vector of currents flowing across edges. If a vertex is an internal vertex, the sum of the currents is zero. This can be stated as $\bar{B}_iI_1 =
\bm{0}_{|V_i|}$. Since we can inject currents into the boundary
vertices, the sum of the currents entering into a boundary vertex is
equal to the injected current. This can be stated as $\bar{B}_bI_1
= I_{0b}$. Therefore, an open graph $\mathcal{G} =
(\mathcal{V},\mathcal{E})$ with the set of boundary vertices
$\mathcal{V}_b$ satisfies KCL for \mbox{$(I_{0b}, I_1) \in
\Lambda_{0b} \times \Lambda_1$} if
\begin{equation}
\label{kcl}
\begin{bmatrix}
\bar{B}_i\\
\bar{B}_b
\end{bmatrix}I_1
=\begin{bmatrix}
\bm{0}_{|\mathcal{V}_i|}\\
I_{0b}
\end{bmatrix}.
\end{equation}
Although KCL, as expressed in~(\ref{kcl}), requires a directed graph, the directions of the currents can be arbitrarily chosen. In other words, they are a modeling choice and not intrinsic to the physics of the problem. 

Kirchoff's Voltage Law states that the sum of voltages along a
cycle is zero. Since the cycle space of a graph
$\mathcal{G}$ with incidence matrix $B$ is $\ker B$, the sum of the voltages along a cycle $K\in \ker B$ can be written as $K^T V^1$ for $V^1 \in \Lambda^1$. Hence, KVL can be written as $K^T V^1=0$ for every $K\in \ker B$ or as $V^1\in (\ker B)^\perp$ with $(\ker B)^\perp$ denoting the orthogonal complement of $\ker B$ with respect to the standard Euclidean inner product. By making the identification $(\ker B)^{\perp} \cong \im B^T$ we can further express KVL as $V^1 = B^T\psi^0$ for some potential $\psi^0 \in \Lambda^0$.

In the remainder of the paper it will be convenient to split the potential $\psi^0\in\Lambda^0$ into its internal components $\psi_i\in\Lambda^0_i$ and its boundary components $\psi_b\in\Lambda^0_b$ resulting in the following version of KVL
\begin{equation}
\label{kvl} V^1 = \bar{B^T}\psi^0 = \bar{B^T_b}\psi^{0b} +
\bar{B^T_i}\psi^{0i}.
\end{equation}
For the rest of the paper, we will abuse notation and denote
$\bar{B}$, $\bar{B_i}$, $\bar{B_b}$ by $B$, $B_i$, $B_b$,
respectively.

\section{Main Result}

Assume that we have a network of electrical components. Each
electrical component has two terminals. The constitutive relation
of an electrical component relates the current
flowing through the electrical component to the voltage across
its terminals. We consider electrical components with constitutive
relations given by:
\begin{equation}
\label{constitutive} \sum_{j=0}^\nu p_{kj} \frac{d^j}{dt^j}I_{1,k}
= \sum_{j=0}^\nu q_{kj} \frac{d^j}{dt^j}V^1_k.
\end{equation}
where $p_{kj},q_{kj} \in \mathbb{R}_{+0}$, $\nu \in \mathbb{N}$ is the highest degree of differentiation, $I_{1,k}$ is the
current flowing through the electrical component $k \in \bar{e}$ and $V^1_k$
is the voltage across the terminals of the electrical component
$k$. The \textit{coefficient vectors} of~(\ref{constitutive}) are
defined as \mbox{$p_k = (p_{k0},\ldots,p_{k\nu}) \in
\mathbb{R}^{\nu+1}_{+0}$} and \mbox{$q_k = (q_{k0},\ldots,q_{k\nu})
\in \mathbb{R}^{\nu+1}_{+0}$}. The \textit{coefficient matrices}
are defined as \mbox{$P = [p_1~|~\ldots~|~p_{e}]$} and \mbox{$Q =
[q_1~|~\ldots~|~q_{e}]$}.

The constitutive relation for a linear ideal resistor is given by $rI_{1,k} = V^1_k$ and can be described by~(\ref{constitutive}) if we take $\nu = 0$ and $\frac{p_{k0}}{q_{k0}} = r$. Similarly, the constitutive relations for inductors and capacitors, $\ell \dot{I}_{1,k} = V^1_k$ and $I_{1,k} = c\dot{V}^1_k$, are described by~(\ref{constitutive}) when we set $\nu = 1$, $p_{k0}=0$, $p_{k1}=\ell$, $q_{k0}=0$, $q_{k1}=1$, and $\nu = 1$, $p_{k0}=0$, $p_{k1}=1$, $q_{k0}=0$, $q_{k1}=c$, respectively. Hence,~(\ref{constitutive}) generalizes the constitutive relations of RLC circuits.

The constant upper bound $\nu$ in~(\ref{constitutive}) is the same for every electrical
component. In other words, the highest degree of differentiation
in~(\ref{constitutive}), which is a measure of the complexity of
the electrical component, is independent of the electrical
component. We can think of an electrical network as a directed
graph in which each edge in the graph represents an electrical
component. In this framework, electrical components relate the
space of trajectories of currents flowing through the edges
$\Lambda_1$ and its dual space, the space of trajectories of
voltages across the edges $\Lambda^1$. In other words, for the
directed graph $\mathcal{G}=(\mathcal{V},\mathcal{E})$, $(I_1,V^1) \in \Lambda_1 \times
\Lambda^1$ satisfies the constitutive
relations~(\ref{constitutive}) if for every edge $e_k \in
\mathcal{E}$, the relationship between $I_{1,k}$ and $V^1_k$ is
given by~(\ref{constitutive}), where $I_{1,k}$ is the
$k^\text{th}$ element of $I_1$ and $V^1_k$ is the $k^\text{th}$
element of $V^1$. For every edge $e_k \in \mathcal{E}$, the
coefficient vectors of~(\ref{constitutive}) are $p_k$ and $q_k$.
For the rest of the paper, we will assume that $p_k \ne
\bm{0}_{\nu+1}$, i.e., no short-circuit edges and $q_k \ne \bm{0}_{\nu+1}$, i.e., no open-circuit edge.

\begin{definition}
A \textit{generalized electrical network} is a five-tuple
\mbox{$\mathcal{N}=(\mathcal{G},\mathcal{V}_b,\nu,P,Q)$}. It
consists of: an open directed graph $\mathcal{G} =
(\mathcal{V},\mathcal{E})$ on which KCL~(\ref{kcl}), KVL~(\ref{kvl}), and constitutive
relations~(\ref{constitutive}) are satisfied; a set of boundary vertices $\mathcal{V}_b \subset \mathcal{V}$; an order of differentiation $\nu$ (the
constant in~(\ref{constitutive})); and the coefficient matrices \mbox{$P,Q \in
\mathbb{R}^{(\nu+1) \times e}$}.
\end{definition}

We adapt the notion of terminal behavior, which was introduced in
\cite{Willems2009,Verriest2010}, to our framework.

\begin{definition}
The terminal behavior $\mathcal{B}_{\mathcal{N}} \subset
\Lambda^0_b \times \Lambda_{0b}$ of a generalized electrical
network \mbox{$\mathcal{N}=(\mathcal{G},\mathcal{V}_b,\nu,P,Q)$}
is the relation defined by: $(\psi^{0b},I_{0b}) \in
\mathcal{B}_{\mathcal{N}}$ iff there exists a $\psi^{0i} \in
\Lambda^{0i}$ such that $\psi^0 = (\psi^{0b},\psi^{0i}) \in
\Lambda^0$ and $I_0 = (I_{0b},\bm{0}_{|\mathcal{V}_e|}) \in
\Lambda_{0b} \times \Lambda_{0i}$ satisfy KCL~(\ref{kcl}), KVL~(\ref{kvl}) and
constitutive relations~(\ref{constitutive}) on $\mathcal{G}$.
\end{definition}

The problem addressed in this note is:
\begin{problem}[Kron Reduction \cite{Dorfler2012}]
\label{mainprob} Given a generalized electrical network
$\mathcal{N}=(\mathcal{G},\mathcal{V}_b,\nu,P,Q)$, when can
we construct another generalized electrical network
$\hat{\mathcal{N}}=(\hat{\mathcal{G}},\mathcal{V}_b,\nu,\hat{P},\hat{Q})$
with $\hat{\mathcal{G}} = (\mathcal{V}_b,\hat{\mathcal{E}})$ and
$\mathcal{B}_\mathcal{N} = \mathcal{B}_{\hat{\mathcal{N}}}$?
\end{problem}

Note that every vertex in the graph $\hat{\mathcal{G}}$ is a
boundary vertex. Therefore Problem~\ref{mainprob} is equivalent to
eliminating all the internal vertices of the generalized
electrical network $\mathcal{N}$ without changing the terminal
behavior and the complexity of the constitutive relations measured by $\nu$. We can now can state our main result.

\begin{theorem}
\label{mainthm} Problem~\ref{mainprob} is solvable for the generalized
electrical network
$\mathcal{N}=(\mathcal{G},\mathcal{V}_b,\nu,P,Q)$ if we have
\begin{equation}
\dm\spn\{p_1, \hdots ,p_e\}=\dm\spn\{q_1, \ldots ,q_e\} =
1,\label{cond}
\end{equation}
where $p_j$, $q_j$ are the coefficient vectors
of~(\ref{constitutive}) for edge $e_j \in \mathcal{E}$, where \mbox{$j \in \{1,\ldots,e\}$}.
\end{theorem}

\textit{Proof.} Assume that
\mbox{$\mathcal{N}=(\mathcal{G},\mathcal{V}_b,\nu,P,Q)$} is a
generalized electrical network satisfying~(\ref{cond}).
condition~(\ref{cond}) states that the vector space $\spn\{p_1,
\ldots,p_e\}$ has dimension one. This implies that the basis for
this vector space consists of a single vector \mbox{$\tilde{p} =
(\tilde{p}_1,\ldots,\tilde{p}_e) \ne \bm{0}_{\nu + 1}$}. Thus for
every $k \in \bar{e}$, there exists a constant $\lambda_k$ such
that $p_k = \lambda_k \tilde{p}$. Since every element of $p_k$
is nonnegative for all $k \in \bar{e}$, we can assume
without loss of generality that every element of $\tilde{p}$ is
nonnegative. Hence, we can assume
$\lambda_k \ge 0$ for all $k \in \bar{e}$. Similarly, the basis
for the vector space $\spn\{q_1, \ldots ,q_e\}$ consists of a
single vector \mbox{$\tilde{q} = (\tilde{q}_1,\ldots,\tilde{q}_e)
\ne \bm{0}_{\nu + 1}$}. From the same reasoning, we can assume $\gamma_k \ge 0$. Replacing $p_k$ and
$q_k$ in~(\ref{constitutive}) with $p_k = \lambda_k \tilde{p}$ and
$q_k = \gamma_k \tilde{q}$, we obtain
\begin{equation}
\lambda_k \sum_{j=0}^\nu \tilde{p}_{j} \frac{d^j}{dt^j}I_{1,k} =
\gamma_k \sum_{j=0}^\nu \tilde{q}_{j}
\frac{d^j}{dt^j}V^1_k,\label{proofedge}
\end{equation}
for every edge $e_k \in \mathcal{E}$. By assumption, we have
\mbox{$p_k \ne \bm{0}_{\nu + 1}$} and $q_k \ne \bm{0}_{\nu + 1}$.
This implies that $\lambda_k \ne 0$ and $\gamma_k \ne 0$. Dividing
(\ref{proofedge}) by $\lambda_k$ for each $k \in \bar{e}$ and
writing equation~(\ref{proofedge}) for every edge $e_k \in
\mathcal{E}$ in vector form, we obtain
\begin{equation}
\sum_{j=0}^\nu \tilde{p}_{j} \frac{d^j}{dt^j}I_1 = \Gamma
\sum_{j=0}^\nu \tilde{q}_{j} \frac{d^j}{dt^j}V^1,\label{proofvec}
\end{equation}
where \mbox{$I_1 = (I_{1,1},\ldots,I_{1,e})$}, $V^1 =
(V^1_{1},\ldots,V^1_{e})$ and $\Gamma$ is a diagonal matrix with
strictly positive diagonal elements. The matrix $\Gamma$ is
defined as $\Gamma_{kk} = \frac{\gamma_k}{\lambda_k}$ for all $k
\in \bar{e}$. From KVL~(\ref{kvl}), we have \mbox{$V^1 =
B^T\psi^0$} for $\psi^0 \in \Lambda_0$. Replacing $V^1$
in~(\ref{proofvec}), we obtain
\begin{equation}
\sum_{j=0}^\nu \tilde{p}_{j} \frac{d^j}{dt^j}I_1 = \Gamma
\sum_{j=0}^\nu \tilde{q}_{j} \frac{d^j}{dt^j}B^T \psi^0 = \Gamma B^T \sum_{j=0}^\nu \tilde{q}_{j}
\frac{d^j}{dt^j}\psi^0.\notag
\end{equation}
and multiplication by $B$ leads to
\begin{align}
B\sum_{j=0}^\nu \tilde{p}_{j} \frac{d^j}{dt^j}I_1 &= B\Gamma B^T
\sum_{j=0}^\nu \tilde{q}_{j} \frac{d^j}{dt^j}
\psi^0 \label{proofb1} \\
\Longleftrightarrow \sum_{j=0}^\nu \tilde{p}_{j}
\frac{d^j}{dt^j}BI_1 &= B\Gamma B^T \sum_{j=0}^\nu \tilde{q}_{j}
\frac{d^j}{dt^j}\psi^0.\label{proofb}
\end{align}
Using the previously defined partitioning of $B$ into $B_i$ and
$B_b$, we obtain the following set of equations
from~(\ref{proofb})
\begin{equation}
\sum_{j=0}^\nu \tilde{p}_{j} \frac{d^j}{dt^j}B_bI_1 = B_b\Gamma B_i^T \sum_{j=0}^\nu \tilde{q}_{j}
\frac{d^j}{dt^j}\psi^{0i} + B_b\Gamma B_b^T
\sum_{j=0}^\nu \tilde{q}_{j} \frac{d^j}{dt^j}\psi^{0b}.\label{proofpart11} 
\end{equation}
\begin{equation}
\sum_{j=0}^\nu \tilde{p}_{j} \frac{d^j}{dt^j}B_iI_1 = B_i\Gamma B_i^T \sum_{j=0}^\nu \tilde{q}_{j}
\frac{d^j}{dt^j}\psi^{0i} + B_i\Gamma B_b^T \sum_{j=0}^\nu
\tilde{q}_{j} \frac{d^j}{dt^j}\psi^{0b}. \label{proofpart12}
\end{equation}
From KCL (\ref{kcl}), we have $B_bI_1 = I_{0b}$ and $B_iI_1 =
\bm{0}_{|\mathcal{V}_i|}$. Replacing $B_bI_1 = I_{0b}$
in~(\ref{proofpart11}) and $B_iI_1 = \bm{0}_{|\mathcal{V}_i|}$
in~(\ref{proofpart12}), we have
\begin{equation}
\sum_{j=0}^\nu \tilde{p}_{j} \frac{d^j}{dt^j}I_{0b} = B_b\Gamma B_i^T \sum_{j=0}^\nu \tilde{q}_{j} \frac{d^j}{dt^j}\psi^{0i} + B_b\Gamma B_b^T
\sum_{j=0}^\nu \tilde{q}_{j} \frac{d^j}{dt^j}\psi^{0b}, \label{proofpart21}
\end{equation}
\begin{equation}
\bm{0}_{|\mathcal{V}_i|} = B_i\Gamma B_i^T \sum_{j=0}^\nu
\tilde{q}_{j} \frac{d^j}{dt^j}\psi^{0i} + B_i\Gamma B_b^T
\sum_{j=0}^\nu \tilde{q}_{j} \frac{d^j}{dt^j}\psi^{0b}.
\label{proofpart22}
\end{equation}
The matrix $B_i\Gamma B^T_i$ is invertible by
Lemma~\ref{lemmakron}. Therefore, we obtain
from the previous equality:
\begin{equation}
\sum_{j=0}^\nu \tilde{q}_{j} \frac{d^j}{dt^j}\psi^{0i} =
-(B_i\Gamma B_i^T)^{-1}B_i\Gamma B_b^T \sum_{j=0}^\nu
\tilde{q}_{j} \frac{d^j}{dt^j}\psi^{0b}. \label{proofpart3}
\end{equation}
Substituting $\sum_{j=0}^\nu \tilde{q}_{j}
\frac{d^j}{dt^j}\psi^{0i}$ into~(\ref{proofpart21}), we obtain
\begin{equation}
\sum_{j=0}^\nu \tilde{p}_{j} \frac{d^j}{dt^j}I_{0b} = \big(B_b \Gamma B^T_b - B_b \Gamma B^T_i (B_i\Gamma
B_i^T)^{-1}B_i\Gamma B_b^T\big) \sum_{j=0}^\nu
\tilde{q}_{j}\frac{d^j}{dt^j}\psi^{0b}.\label{terminalG}
\end{equation}
Smoothness of $\psi^{0b}$ implies that the left hand side
of~(\ref{terminalG}) and the left hand side of~(\ref{proofpart3})
are continuous functions. Since $\tilde{p} \ne \bm{0}_{\nu + 1}$
for any $\psi^{0b} \in \Lambda^0_b$, there exists a unique
\mbox{$I_{0b} \in \Lambda_{0b}$} that satisfies~(\ref{terminalG})
and a unique $\psi^{0i}$ that satisfies~(\ref{proofpart3}).
Therefore, if $I_{0b}$ and $\psi_{0b}$ satisfy~(\ref{terminalG}),
then there exists a unique $\psi^{0i} \in \Lambda^{0i}$ such that
$\psi^0 = (\psi^{0b},\psi^{0i}) \in \Lambda^0$ and $I_0 =
(I_{0b},\bm{0}_{|\mathcal{V}_e}|) \in \Lambda_0$ satisfy
KCL~(\ref{kcl}), KVL~(\ref{kvl}) and constitutive
relations~(\ref{constitutive}). Thus $(\psi_b,I_b) \in
\mathcal{B}_{\mathcal{N}}$ iff $I_{0b}$ and $\psi_{0b}$
satisfy~(\ref{terminalG}). We now want to construct a generalized
electrical network
\mbox{$\hat{\mathcal{N}}=(\hat{\mathcal{G}},\mathcal{V}_b,\nu,\hat{P},\hat{Q})$}
with $\hat{\mathcal{G}} = (\mathcal{V}_b,\hat{\mathcal{E}})$ and
$\mathcal{B}_\mathcal{N} = \mathcal{B}_{\hat{\mathcal{N}}}$. From
Lemma~\ref{lemmakron}, there exists a graph $\hat{\mathcal{G}} =
(\mathcal{V}_b,\hat{\mathcal{E}})$ with incidence matrix $\hat{B}$
and a diagonal matrix $\hat{\Gamma}$ with strictly positive
diagonal elements such that
\begin{equation}
\hat{B}\hat{\Gamma}\hat{B}^T = B_b\Gamma B^T_b - B_b\Gamma B^T_i
(B_i\Gamma B^T_i)^{-1} B_i\Gamma B^T_b. \label{schur}
\end{equation}
We construct a generalized electrical network $\hat{\mathcal{N}}$
from the directed graph $\hat{\mathcal{G}}$ by defining the
constitutive relations on $\hat{\mathcal{G}}$ as
\begin{equation}
\sum_{j=0}^\nu \tilde{p}_{j} \frac{d^j}{dt^j}\hat{I}_1 =
\hat{\Gamma} \sum_{j=0}^\nu \tilde{q}_{j}
\frac{d^j}{dt^j}\hat{V}^1.\label{terminalGhat}
\end{equation}
Multiplying both sides of~(\ref{terminalGhat}) by $\hat{B}$ and
using~(\ref{kcl}),~(\ref{kvl}),~(\ref{schur}); we
obtain~(\ref{terminalG}) from~(\ref{terminalGhat}). Therefore we
can construct a generalized electrical circuit $\hat{\mathcal{N}}$
that has the same terminal behavior as $\mathcal{N}$.\hfill$\Box$

We emphasize that the reduction process detailed in the proof of Theorem~\ref{mainthm} is performed in the time domain and requires no steady state assumptions. We start with the generalized electrical network $\mathcal{N}$ that describes the relation between voltages and currents in the time domain and we construct the unique Kron reduced generalized electrical network $\hat{\mathcal{N}}$ that has the same terminal behavior also described in the time domain by the constitutive relations~(\ref{terminalGhat}). In the proof we assumed that no currents are injected at the internal vertices. However, the proof smoothly extends to the case where there are injected currents. We shall return to this observation, in the context of power networks, in Section~\ref{Application}.

The following example illustrates that Condition~(\ref{cond}) in
Theorem~\ref{mainthm} is not necessary.

\begin{example} Consider the generalized electrical network
\mbox{$\mathcal{N}=(\mathcal{G},\mathcal{V}_b,\nu,P,Q)$} with
$\mathcal{G}$ given below:
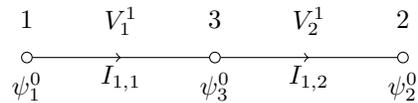
\begin{center}
\begin{tikzpicture}[every edge/.style={draw,postaction={decorate,decoration={markings,mark=at position 0.5 with {\arrow{>}}}}}]
    \node[vertex] (1) at (1,0) [label=below:$\psi^0_1$]{};
    \node[vertex] (3) at (3.5,0) [label=below:$\psi^0_3$]{};
    \node[vertex] (2) at (6,0) [label=below:$\psi^0_2$]{};
    \node[emptynode] at (2.25,-0.75) [label=above:$I_{1,1}$] {};
    \node[emptynode] at (4.75,-0.75) [label=above:$I_{1,2}$] {};
    \node[emptynode] at (2.25,0) [label=above:$V^1_{1}$] {};
    \node[emptynode] at (4.75,0) [label=above:$V^1_{2}$] {};
    \node[emptynode] at (1,0.1) [label=above:$1$] {};
    \node[emptynode] at (3.5,0.1) [label=above:$3$] {};
    \node[emptynode] at (6,0.1) [label=above:$2$] {};
    \path
    (1) edge (3)
    (3) edge (2);
\end{tikzpicture}
\end{center}
The constitutive relations for the edges $e_1 = (1,3)$ and $e_2 =
(3,2)$ are $I_{1,1} = V^1_1$ and $\frac{d}{dt}I_{1,2} = V^1_2$,
respectively. Note that $\nu = 1$. From KCL~(\ref{kcl}), we have
$I_{0b,1} = I_{1,1}$, \mbox{$I_{1,1} = -I_{1,2}$} and $I_{1,2} =
-I_{0b,2}$. From KVL~(\ref{kvl}), we have \mbox{$V^1_1 = \psi^0_3
- \psi^0_1$} and $V^1_2 = \psi^0_2 - \psi^0_3$. Replacing
$I_{1,1}$, $I_{1,2}$, $V^1_1$ and $V^1_2$ in the constitutive
relations, we obtain
\begin{align}
I_{0b,1} = -I_{0b,2} = \psi^0_3 - \psi^0_1, \notag \\
\frac{d}{dt}I_{0b,1} = -\frac{d}{dt}I_{0b,2} = \psi^0_2 - \psi^0_3
\notag.
\end{align}
Combining these equations, we obtain
\begin{equation}
I_{0b,1} + \frac{d}{dt}I_{0b,1} = -I_{0b,2} -\frac{d}{dt}I_{0b,2}
= \psi^0_2 - \psi^0_1.\label{term1}
\end{equation}
There exists a $\psi^0_3$ such that $(\psi^0_1,\psi^0_2,\psi^0_3)$
and $(I_{0b,1},I_{0b,2})$ satisfy KCL~(\ref{kcl}), KVL~(\ref{kvl})
and the constitutive relations if and only if
$(\psi^0_1,\psi^0_2)$ and $(I_{0b,1},I_{0b,2})$
satisfy~(\ref{term1}). Therefore
$(\psi^0_1,\psi^0_2,I_{0b,1},I_{0b,2}) \in
\mathcal{B}_{\mathcal{N}}$ if and only if
$(\psi^0_1,\psi^0_2,I_{0b,1},I_{0b,2})$ satisfy~(\ref{term1}). We
now construct a generalized electrical network
\mbox{$\hat{\mathcal{N}}=(\hat{\mathcal{G}},\mathcal{V}_b,1,P,Q)$}
such that $\mathcal{B}_{\hat{\mathcal{N}}} =
\mathcal{B}_{\mathcal{N}}$. The directed graph $\hat{\mathcal{G}}$
is given below.
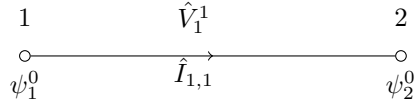
\begin{center}
\begin{tikzpicture}[every edge/.style={draw,postaction={decorate,decoration={markings,mark=at position 0.5 with {\arrow{>}}}}}]
    \node[vertex] (1) at (1,0) [label=below:$\psi^0_1$]{};
    \node[vertex] (2) at (6,0) [label=below:$\psi^0_2$]{};
    \node[emptynode] at (3.25,-0.75) [label=above:$\hat{I}_{1,1}$] {};
    \node[emptynode] at (3.25,0) [label=above:$\hat{V}^1_{1}$] {};
    \node[emptynode] at (1,0.1) [label=above:$1$] {};
    \node[emptynode] at (6,0.1) [label=above:$2$] {};
    \path
    (1) edge (2);
\end{tikzpicture}
\end{center}
We pick the constitutive relation of the single edge in
$\hat{\mathcal{N}}$ as
$$
\hat{I}_{1,1} + \frac{d}{dt}{\hat{I}}_{1,1} = \hat{V}^1_{1}.
$$
From KCL~(\ref{kcl}), we have $I_{0b,1} = -I_{0b,2} =
\hat{I}_{1,1}$. From KVL~(\ref{kvl}), we have $\hat{V}^1_1 =
\psi^0_2 - \psi^0_1$. Replacing $\hat{I}_{1,1}$ and $\hat{V}^1_1$
in the constitutive relations, we recover~(\ref{term1}). Therefore
$(\psi^0_1,\psi^0_2,I_{0b,1},I_{0b,2}) \in
\mathcal{B}_{\hat{\mathcal{N}}}$ if and only if
$(\psi^0_1,\psi^0_2,I_{0b,1},I_{0b,2})$ satisfy~(\ref{term1}).
This implies that $\mathcal{B}_{\mathcal{N}} =
\mathcal{B}_{\hat{\mathcal{N}}}$. Hence Problem~\ref{mainprob} is
solvable. However $p_1 = (1,0)$ and $p_2 = (0,1)$. Therefore $\dm
\spn \{p_1,p_2\} = 2$ and condition~(\ref{cond}) does not hold.
\end{example}

We now provide an example for which condition~(\ref{cond}) in Theorem~\ref{mainthm} fails to hold and Kron reduction is impossible.

\begin{example} Consider the generalized electrical network in the previous example. If we pick the constitutive relations for the edges of $\mathcal{N}$ as $\frac{d}{dt}I_{1,1} = V^1_1$ and $I_{1,2} =\frac{d}{dt}V^1_2$, condition~(\ref{cond}) does not hold. Nevertheless, we can compute the constitutive relation for the Kron reduced network as
\begin{equation}
\label{example7}
\frac{d}{dt}(V^1_1 + V^1_2) = \frac{d}{dt}(\psi^0_2-\psi^0_1) = \frac{d^2}{dt^2}I_{1,1} + I_{1,2}.
\end{equation}
Observe that for the original network we have $\nu = 1$ while for the reduced network we have $\nu=2$. Hence Problem~\ref{mainprob} is not solvable in this case. 
\end{example}

\section{Application to RLC Circuits and Power Networks}
\label{Application}

\textit{Every} RLC circuit can be modeled as a generalized
electrical network by taking the electric components to be combinations of
resistors, inductors, or capacitors. When all the circuit elements are resistors we speak of a purely resistive circuit. Purely inductive and purely capacitive circuits can be defined similarly. It is shown in \cite{Arjan2010} that Problem~\ref{mainprob} is solvable for purely resistive, inductive, or capacitive circuits. The same result can be obtained by the following corollary of Theorem~\ref{mainthm}.

\begin{corollary}
\label{corpure} Problem~\ref{mainprob} is solvable for the
generalized electrical network $\mathcal{N}$ if $\mathcal{N}$ is a
purely resistive, purely inductive or purely capacitive circuit.
\end{corollary}

\textit{Proof. }We will prove the corollary for purely resistive
circuits. The proofs for purely inductive and purely capacitive
circuits are very similar. In a purely resistive circuit,
$\bm{p}_i = (r_i,0)$, and $\bm{q}_i = (1,0)$, where $r_i \in
\mathbb{R}^e_+$ is the resistance of the edge i. Hence, condition~(\ref{cond}) holds.
The result follows from Theorem~\ref{mainthm}.\hfill$\Box$

One particular example of generalized electrical networks is
\textit{homogeneous} RL circuits \cite{Caliskan2012}. Every edge
of an homogeneous RL circuit is a series connection of a resistor
and an inductor. The term homogeneous comes from the fact that for
every two edge $e_i,e_j \in \mathcal{E}$ with resistance values
$r_i,r_j$ and inductor values $\ell_i,\ell_j$, we have
$\frac{r_i}{r_j} = \frac{\ell_i}{\ell_j}$. In order to represent
RL circuits, it is enough to set $\nu = 1$ and
$(p_{i1},p_{i2},q_{i1},q_{i2}) = (r_i,\ell_i,1,0)$
in~(\ref{constitutive}) for all $i \in \bar{e}$, where $r_i$ is
the resistance value of the resistive component of edge $i$ and
$\ell_i$ is the inductance value of the inductive component of
edge $i$. Note that homogeneity implies that there exists a
constant $c \in \mathbb{R}$, $c>0$ such that $\bm{p}_i = c
\bm{p}_j$ for every $i,j \in \bar{e}$. Therefore condition~(\ref{cond})
 holds
and we can recover Theorem 4.4 in \cite{Caliskan2012} as a
corollary of Theorem~\ref{mainthm}. The concept of homogeneity can
be generalized to RLC circuits. A \textit{homogeneous} RLC circuit
is an electrical circuit such that every edge is a series
combination of a resistor, an inductor and a capacitor with the
following condition: for every two edges $e_i,e_j \in \mathcal{E}$
with resistance values $r_i,r_j$, inductor values $\ell_i,\ell_j$,
and capacitor values $c_i,c_j$; we have $\frac{r_i}{r_j} =
\frac{\ell_i}{\ell_j} = \frac{c_i}{c_j}$. \textit{Homogeneous}
$RC$ circuits and \textit{homogeneous} $LC$ circuits can be
defined in a similar fashion. The previous discussion is
summarized in the next result.
\begin{corollary} \label{corhom} Problem~\ref{mainprob} is solvable
for homogeneous RLC, RL, RC or LC circuits.
\end{corollary}

There are various transmission line models available in the literature \cite{DeshpandeBook}, \cite{SauerBook} \cite{BergenBook}. If the transmission line is
relatively short (less than 60 kilometers \cite{DeshpandeBook}, or
50 miles \cite{BergenBook}), the transmission line can be described by the short line approximation. In the short line approximation the line is modeled as a
series connection of a resistor and an inductor \cite{BergenBook}. 
Hence every network of short transmission lines can be modeled as
a generalized electrical network with $\nu = 1$, $\bm{p}_i =
(r_i,\ell_i)$ and $\bm{q}_i = (1,0)$ in~(\ref{constitutive}),
where $r_i$ is the resistance value of the resistive component of
edge $i$ and $\ell_i$ is the inductance value of the inductive
component of edge $i$.  Moreover, if the network is a homogeneous
RL circuit, it follows from Corollary~\ref{corhom} that we can
perform Kron reduction. To guarantee that the homogeneity condition only depends on the transmission lines, we model the loads as (possibly nonlinear) current injections at the internal vertices. As discussed after Theorem~\ref{mainthm}, in this case Kron reduction is still possible and the aggregated effect of the loads will appear as (possibly nonlinear) current injections at the boundary vertices of the reduced network.

Different types of transmission line models are suitable for the analysis of different types of transients \cite{Dommel97}. The short line approximation is used to analyze the electromechanical transients \cite{PavellaBook}, \cite{Dommel97}.  One can argue that the assumption that allows us to use the short line approximation (sinusoidal voltages and currents) to study electromechanical transients also allows us to use phasors. However, in our framework we do not need to assume sinusoidal waveforms \textit{per se}. As long as the short line approximation can accurately describe the transients in consideration, we can use the reduced model for transient analysis.

\end{document}